# Temperature coefficient of Silicon based carrier selective solar cells

Nithin Chatterji, Aldrin Antony, and Pradeep R. Nair

*Abstract* – Carrier Selective (CS) Silicon solar cells are increasingly explored as a low cost alternative to PN junction Silicon solar cells. While the recent trends on power conversion efficiency are encouraging, the temperature coefficient and hence the power output under elevated temperatures are not well explored for such solar cells. Here, we address this issue through detailed numerical simulations to explore the influence of interface and material parameters on the temperature coefficient. Our results indicate that irrespective of the interface quality, the temperature coefficient of CS solar cells improves with an increase in band discontinuities. Interestingly, contrary to the trends related to efficiency, our results indicate that the temperature coefficient of CS solar cells is more critically affected by the interface quality of the minority carrier extraction layer than the majority carrier extraction layer. These insights have important implications towards the choice of optimal material and processing conditions for Si based CS solar cells.

## I. Introduction

Si based CS solar cells are considered as a good candidate to challenge the market dominance of conventional PN junction based solar cells. The perceived competitiveness stems from a few advantages like – (a) large band gap materials used as CS layers provides carrier selectivity through appropriate band offsets[1][2], (b) it reduces the parasitic absorption as compared to the heavily doped emitter layer in PN junction solar cells or doped a-Si in HIT solar cells[3], (c) the intrinsic doping density of many CS materials[4]–[6] could be significant enough to eliminate the need for any additional intentional doping process, and (d) possibility of low temperature deposition processes for CS layers[6]–[11]. These aspects could lead to reduced thermal budget and hence lower the cost of fabrication - which motivates the significant recent research interest in CS based solar cells. Accordingly, different materials such as $TiO_2$[4], [5], [12], a:Si[3], poly-Si[13], $LiF_x$[7], $KF_x$[8], PEDOT:PSS[6], $MoO_x$[9]–[11], [14],$V_2O_5$[11], and $WO_3$[11] have been extensively studied. Similarly, there are numerous simulations and analytical studies to understand the working of these types of solar cells [15], [16][17]. However, all these studies were at standard test conditions which could be significantly different from the actual conditions at the place of deployment. The eventual power output from a panel at actual conditions could be significantly different from that at STC due to the strong dependence of efficiency on temperature. In this regard, the temperature coefficient of a solar cell is often treated as a critical parameter to ascertain and compare the performance of various technologies. Indeed, the temperature coefficient of PN junction[18], [19] and HIT solar cells[20] are well explored. However, such a study has not been reported yet on CS solar cells. Since diverse materials are being investigated as CS layers, a-priori knowledge of temperature coefficient as a function of material parameters would be immensely beneficial.

In this manuscript, the temperature coefficient for CS based solar cells is established through detailed modeling and is then compared with the well-established HIT and PERC solar cell technologies. For this, we first develop an analytical model to predict the functional dependence of temperature coefficient on important parameters like band discontinuity and interface quality (Section II). These predictions are then validated through detailed numerical simulations (Section III). Our results indicate that the temperature coefficient is limited by $V_{oc}$ for smaller band discontinuities and it follows the trend of FF for larger magnitudes of band discontinuity. Additionally, unlike efficiency[17], temperature coefficient of CS based solar cells is dominated by the interface quality of the minority carrier collection junction. Below we first develop an analytical model to predict the temperature coefficient.

## II. Analytical Model

Figure 1 shows band level alignments of Si heterojunction solar cell with CS layers (the material parameters are provided in the appendix). ESL and HSL in the figure denote the electron and hole selective layers, respectively. The large valence band offset at ESL/Si interface blocks the transport of holes from Si to ESL. However, the smaller conduction band offset at the same interface aids the transport of electrons to ESL. Note that a positive value for band offset indicates that the photo-generated carriers need to overcome a barrier to reach the corresponding transport layer, where as a negative value for the offset indicates that the carrier injection from transport layer to silicon is limited by a potential barrier. Further there could be imperfections in the interface in the form of traps. Similarly, we assume that the Si/HSL interface to be perfect electron blocking with a smaller valence band offset that aids hole collection. Here we explicitly consider the temperature coefficient variations of such solar cells as a

The authors are with Indian Institute of Technology Bombay, Mumbai 400076, India (email: nithin@ee.iitb.ac.in; aldrinantony@iitb.ac.in; prnair@ee.iitb.ac.in)

function of transport barrier and interface trap density. As such, temperature coefficient could also be influenced by parameters like the effective doping and thickness of carrier selective layers, nature of metal or TCO contact with the carrier selective layers, etc. With the aim of developing a coherent description of the various effects, here we make a few simplifying assumptions – (a) the contact layers are assumed to be doped, (b) the metal or TCO contact with selective layers are assumed to be ohmic in nature, (c) over the barrier transport is assumed as the dominant transport mechanism at Si/CS layer interface, and (d) uniform density of traps at Si/CS layer interface. With these assumptions we first develop an analytical model to predict the device performance. Later, results from detailed numerical simulations (self-consistent solution of Poisson and carrier continuity equations) are provided to further refine analytical predictions. The parameters used in this study are listed in appendix A.

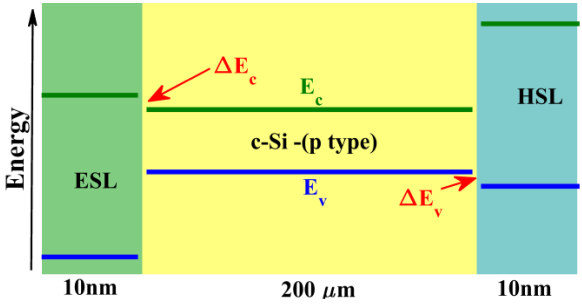

**Fig. 1**: Energy level alignments of a Si based carrier selective solar cell. Refer Table 1 for simulation parameters.

The temperature coefficient of the efficiency is defined as,

$$TC\eta = \frac{1}{\eta(STC)}\frac{\partial \eta}{\partial T},\qquad(1)$$

where STC denotes the standard test conditions, which assumes T=298K in this manuscript. The efficiency of a solar cell, in turn, depends on the open circuit potential ($V_{oc}$), short circuit current ($J_{sc}$), and the fill factor (FF)[21]. Accordingly, a first order estimate for $TC\eta$ can be obtained in terms of the individual temperature coefficients as given below

$$\frac{1}{\eta(STC)}\frac{\partial \eta}{\partial T} = \frac{1}{V_{oc}(STC)}\frac{\partial V_{oc}}{\partial T} + \frac{1}{J_{sc}(STC)}\frac{\partial J_{sc}}{\partial T} + \frac{1}{FF(STC)}\frac{\partial FF}{\partial T}.\qquad(2)$$

Here the terms in the RHS denote temperature coefficients of $V_{oc}$ ($TCV_{oc}$), $J_{sc}$ ($TCJ_{sc}$), and FF (TCFF), respectively. Note that each of these temperature coefficients is defined against the respective parameter at STC. As such, good estimates for the temperature coefficient of efficiency can be obtained through detailed knowledge about the temperature variation of parameters like $V_{oc}$, $J_{sc}$, and FF. With this aim, and to develop quantitative insights, we first explore the effect ESL/Si interface parameters on $TC\eta$ with ideal conditions assumed for Si/HSL interface (i.e., zero band offset for holes, perfect electron blocking, and lack of any interface trap states).

a ) Effect of $\Delta E_c$ on $\frac{1}{V_{oc}}\frac{\partial V_{oc}}{\partial T}$: It is well known that the temperature coefficient of solar cell is dominated by the temperature sensitivity of $V_{oc}$[22]. For a solar cell the maximum achievable $V_{oc}$ is dictated by the detailed balance of carrier generation with various recombination mechanisms. The asymptotic limits of $V_{oc}$ for a CS solar cell in the presence of interface traps was recently shown as [17]

$$V_{oc,max} = \frac{kT}{q}\ln\left(\frac{G\,\tau_b N_A}{n_i^2}\right),\qquad(3)$$

$$V_{oc,min} = \frac{2kT}{q}\ln\left(\frac{2Gw_{Si}}{c_{ns}D_{it}E_{g,Si}n_i}\right).\qquad(4)$$

Here, $V_{oc,max}$ and $V_{oc,min}$ are the maximum and minimum values of $V_{oc}$ respectively, $G$ is the carrier generation rate in Silicon, $\tau_b$ is the effective bulk life time in Silicon (the combined effects of SRH, radiative, and Auger recombinations), $N_A$ is the acceptor doping in Silicon, $n_i$ is the intrinsic carrier concentration in Silicon, $D_{it}$ is the interface defect density at the ESL/Si interface and $E_{g,Si}$ is the bandgap of Silicon. Note that $V_{oc,max}$ is influenced only by the bulk lifetime while $V_{oc,min}$ is entirely dictated by the interface recombination. Although eq. (4) considers interface recombination at Si/ESL junction only, it could be easily modified to account for the recombination in Si/HSL junction as well[17].

Equations (3) and (4) allow us to estimate the asymptotic limits for temperature coefficient of $V_{oc}$. Accordingly, the maximum and minimum values of $TCV_{oc}$ is determined as $-0.19\%/°C$ and $-0.37\%/°C$, respectively for typical values of one sun generation with the $D_{it} = 10^{12}cm^{-2}eV^{-1}$ and $\tau_b = 1ms$. The values for radiative and auger coefficients are given in Appendix A. The dominant temperature dependence of both the limits is through $n_i^{-2}$ which has an exponential relation with temperature ($n_i^2 \sim e^{-E_g/KT}$). Hence the temperature coefficient, $\frac{1}{V_{oc}(STC)}\frac{\partial V_{oc}}{\partial T}$, is negative. The smaller the magnitude of temperature coefficient for $V_{oc}$, the better is its performance at high temperature.

Although the above discussion allows us to estimate the asymptotic limits of $TCV_{oc}$, it still lacks quantitative information of the functional dependence of $TCV_{oc}$ on various parameters like band offset $\Delta E_c$, interface state density $D_{it}$, etc. To obtain the same, we rely on the modified analytical model for $V_{oc}$ reported in ref.[17]. The same model indicates that $V_{oc}$ varies almost symmetrically with $\Delta E_c$, with the minimum at $\Delta E_c=0$. This is due to the fact that interface recombination maximizes under such conditions. The model also indicates that $V_{oc}$ improves for large $\Delta E_c$ due to the reduction in interface recombination. Further, the same model anticipates the variation with parameters like $D_{it}$, $N_A$, $\varepsilon_{ESL}$, etc.

Figure 2(a) shows the variation of $V_{oc}$ with temperature as function of band discontinuity $\Delta E_c$ between ESL and Si with $D_{it} = 10^{12} cm^{-2} eV^{-1}$. As expected, the $V_{oc}$ decreases with temperature due to its strong dependence of $n_i$.

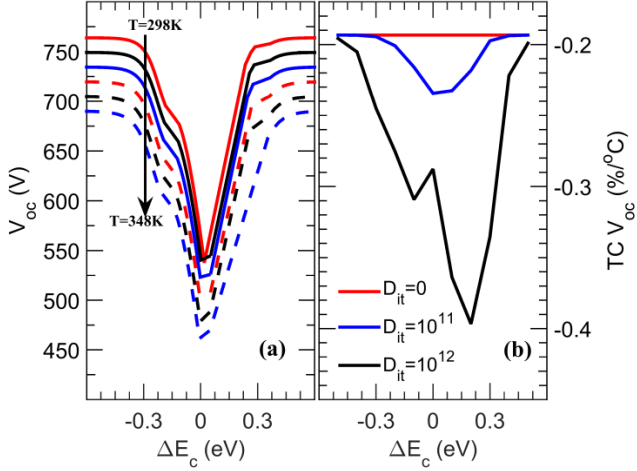

**Fig. 2**: Variation of $V_{oc}$ with $\Delta E_c$, temperature and $D_{it}$. Part (a) shows the variation of $V_{oc}$ with temperature for $D_{it} = 10^{12} cm^{-2} eV^{-1}$. Part (b) shows the variation of temperature coefficient with $D_{it}$.

Figure 2(b) shows the variation in the temperature coefficient of $V_{oc}$ with interface trap density, $D_{it}$, and $\Delta E_c$. In general, the temperature coefficient degrades with an increase in $D_{it}$ as the contribution due to interface recombination becomes significant. However, for large $\Delta E_c$, the interface recombination decreases due to enhanced field effect passivation of traps which results in an increase in both $V_{oc}$ and $TCV_{oc}$.

b) Effect of $\Delta E_c$ on $J_{sc}$ and FF: $J_{sc}$ depends on the number of absorbed photons in c-Si bulk and hence increases with temperature as Si bandgap decreases with temperature. This leads to a positive temperature coefficient of 0.0156%/°C which is very small compared to the $TCV_{oc}$ reported in previous section. If we further account for the fact that the parasitic absorption in the ESL also increases with temperature, then the effect of temperature on $J_{sc}$ is expected to decrease further. As this effect is insignificant relative to that due to $V_{oc}$, we neglect the effect of band gap variation on $J_{sc}$ and assume that the effective photo-carrier generation rate is invariant with temperature.

While the net carrier generation rate could be rather insensitive to temperature variation, the charge collection efficiency could show strong temperature dependence and hence influence $J_{sc}$. Over the barrier transport of carriers across the ESL/Si junction is a dominant mechanism for collection of photo-generated carriers which might introduce additional such temperature dependence on $J_{sc}$. At short circuit conditions, the interface recombination is negligible and the interface carrier density is given as $n_s \sim \frac{n_i^2}{N_A} e^{+\frac{q\psi_{Si}}{kT}}$, where $\psi_{Si}$ is the band bending in Si at ESL/Si interface.

Hence, the current through the ESL, which is $J_{sc}$, is given as [17]

$$J_{sc} \propto n_s e^{-\frac{\Delta E_c}{kT}}. \tag{5}$$

The above equation can be conveniently interpreted in terms of the supply of electrons ($n_s$ on the RHS) and the factor $e^{-\frac{\Delta E_c}{kT}}$ denotes the probability of carriers crossing the barrier due to band offset (i.e., for $\Delta E_c > 0$. Otherwise, the current will be limited by the diffusion of electrons from bulk to Si/ESL interface).

It has been recently shown that the band bending in Si ($\psi_{Si}$) at the Si/ETL interface increases with increase in $\Delta E_c$ [17]. Accordingly, the $n_s$ increases with increase in $\Delta E_c$ and compensates for the reduction in the probability of crossing the barrier (see eq. 5) and hence $J_{sc}$ remains invariant till strong inversion happens[17]. So, the temperature coefficient will be negligible for small $\Delta E_c$. However, for large $\Delta E_c$, strong inversion conditions arise at Si interface and hence the $n_s$ will no longer increase with $\Delta E_c$. This results in a decrease in $J_{sc}$ for large $\Delta E_c$, as predicted by eq. 5. However, for such cases, the probability of carriers crossing the barrier increases with temperature and as a result $J_{sc}$ also improves with temperature (see eq. 5). Hence, the temperature coefficient for large band offsets will be positive.

FF, which is a measure of collection efficiency at maximum power point, was shown in ref. [17] to follow the trends of $V_{oc}$ for small $\Delta E_c$ and $J_{sc}$ for large $\Delta E_c$. Accordingly, for small $\Delta E_c$ the temperature coefficient of FF will be proportional to the temperature coefficient of $V_{oc}$. However, for large $\Delta E_c$, the FF follows the trend of current at maximum power point similar to that of $J_{sc}$. Accordingly, the FF improves with temperature due to the increase in probability of carriers crossing the barrier, and hence its temperature coefficient will be positive.

Finally, the temperature coefficient of efficiency which is the sum of the temperature coefficients of $V_{oc}$, $J_{sc}$ and FF is significantly affected by the $V_{oc}$ for lower band discontinuities. For these cases, the FF too follows $V_{oc}$ trends. For larger band discontinuity ($> 0.4 eV$) the temperature coefficient is dictated by FF, as over the barrier transport of carriers at maximum power point increases with temperature. The analysis in this section predicts that temperature coefficient of efficiency improves with increase in $\Delta E_c$. Interestingly, the temperature coefficient might become positive for large $\Delta E_c$, however, the overall efficiency could still be lower than the corresponding values for small $\Delta E_c$. Therefore, there is an optimal band offset considering both the tradeoff between efficiency and temperature coefficient, which will be explored in detail using numerical simulations.

### III. NUMERICAL SIMULATIONS

In the previous section, we used the analytical model to explore the variation of temperature coefficient as a function of band discontinuity and interface quality. The model predicts that TCη closely follows the features of $TCV_{oc}$. In this section we discuss the results of detailed numerical simulations. Current-Voltage characteristics of the modeled device was obtained through self-consistent solution of Poisson and drift diffusion equations [23]. From the obtained characteristics, temperature coefficient of $V_{oc}, J_{sc}$, FF, and efficiency were estimated. Note the TCη is estimated directly through eq. (1) and as such the numerical simulations do not rely on the simplifying assumptions made in the derivation of the analytical model (for example, validity of eq. (2), and also see the discussion on FF in previous section) and hence can be used as a test bed for analytical predictions. As before, initially we study the effect of conduction band offset $\Delta E_c$ at the ESL/Si interface on the temperature coefficient while keeping Si/HSL interface perfect ($\Delta E_v = 0eV$ and $D_{it} = 0$). The interface trap density at the ESL/c-Si interface is assumed to be uniform as in the analytical model. Later the effect of band offsets at Si/HSL interface and the combined effect of both the offsets are also discussed. The list of parameters used in simulations is given in Appendix A.

The influence of band offset on the interface recombination is highlighted in Fig. 3. Here we consider two cases of $\Delta E_c$ and the subsequent effect on interface recombination. A comparison of parts (a) and (b) of Fig. 3 indicates that the band bending in Si near Si/ESL interface increases with $\Delta E_c$. Accordingly, the corresponding minority carrier density decreases with an increase in band offset (see parts (c) and (d) of Fig. 3).As a result, the interface recombination reduces and hence $V_{oc}$ increases as $\Delta E_c$ increases. Further, the increase in minority carrier density with temperature is also influenced by $\Delta E_c$. The increase in minority carrier density with temperature leads to an increase in interface recombination and hence reduces $V_{oc}$.

respect to the corresponding value (same $\Delta E_c$) at STC conditions and then multiplying by 100. Figure 4a shows that the slope for the normalized $V_{oc}$ is more for $\Delta E_c = 0$ and that the slope decreases as the magnitude of $\Delta E_c$ increases. This is expected from the analytical model on $V_{oc}$ and also supported by the simulation results provided in Fig. 3. For small $\Delta E_c$, the $V_{oc}$ is dominated by interface recombination whose contribution decreases as $\Delta E_c$ increases. Accordingly, the $V_{oc}$ degradation decreases as $\Delta E_c$ increases. Normalized $J_{sc}$ (Fig. 4b) remains invariant with $\Delta E_c$ for the temperature range under consideration due to (a) the assumption of a constant generation rate, and (b) the fact that over the barrier transport of carriers is yet to be significantly influenced by the band offsets – these arguments are consistent with those provided in the analytical section. Normalized FF (Fig. 4c) follows the trends of $V_{oc}$ for lower values of $\Delta E_c$. For $\Delta E_c = 0.5eV$, the FF increases with temperature initially and then saturates. This rather surprising result could be understood as follows: For large $\Delta E_c$, the current at maximum power point conditions at STC is dominated by over the barrier transport limitations due to $\Delta E_c$ band offset. An increase in temperature results in larger over the barrier transport, which leads to better charge collection efficiency and hence the FF. For still higher temperatures the increase in collection efficiency is counter balanced by the decrease in $V_{oc}$ and this result in the saturation of FF (see Fig. 4c). While the trends indicate that it is beneficial to have large $\Delta E_c$ from the perspective of temperature coefficient, we stress that the efficiency is also a function of $\Delta E_c$ (see inset in Fig. 4d for effect of $\Delta E_c$ on efficiency at STC with $D_{it} = 10^{12} cm^{-2} eV^{-1}$). Accordingly, maximum power output at higher temperature could still be given by an optimum $\Delta E_c$.

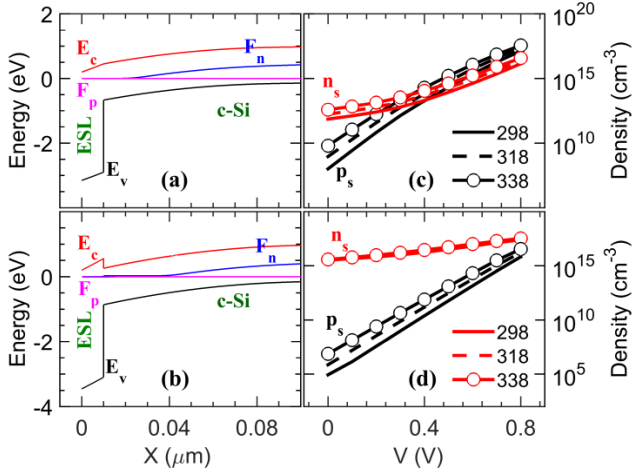

**Fig. 3**: The effect of $\Delta E_c$ on band bending and carrier densities at ESL/c-Si interface with $D_{it} = 10^{12} cm^{-2} eV^{-1}$ (numerical simulations, under illumination). Parts (a) and (b) show the energy band diagram for $\Delta E_c = 0eV$ and $\Delta E_c = 0.3eV$, respectively, at short circuit conditions. Parts (c, d) show the variation in interface carrier densities $n_s$, $p_s$ with bias and temperature.

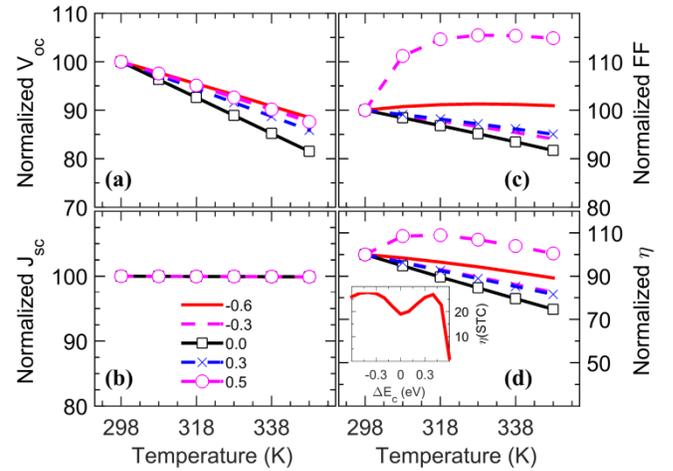

**Fig. 4**: The effect of the $\Delta E_c$ and temperature on normalized performance metrics- (a) $V_{oc}$, (b) $J_{sc}$, (c) FF, and (d) efficiency of CS Si solar cells for $D_{it} = 10^{12} cm^{-2} eV^{-1}$. The labels represent $\Delta E_c = -0.6eV, -0.3eV, 0eV, 0.3eV$, 0 and 0.5eV. The inset in (d) shows the effect of $\Delta E_c$ on efficiency with $D_{it} = 10^{12} cm^{-2} eV^{-1}$.

Figure 4 shows the variation of normalized performance metrics with temperature and $\Delta E_c$. Normalization is done with

Figure 5 shows the temperature coefficient for all performance parameters as a function of $\Delta E_c$ and $D_{it}$. The trends in Fig. 5a are broadly similar to that in the analytical model for temperature coefficient of $V_{oc}$ (see Fig.2b). As explained already in the analytical section (see eq.4 and Fig. 3), $V_{oc}$ and its temperature coefficient is indeed limited by the interface traps when $\Delta E_c$ is low. Further, for larger values of $\Delta E_c$, $V_{oc}$ improves due to the field effect passivation of the interface traps thus improving its temperature coefficient as well. As explained before, this improvement in passivation is due to the increase in band bending with increase in $\Delta E_c$ between ESL/Si and the corresponding decrease in the minority carrier concentration at the interface.

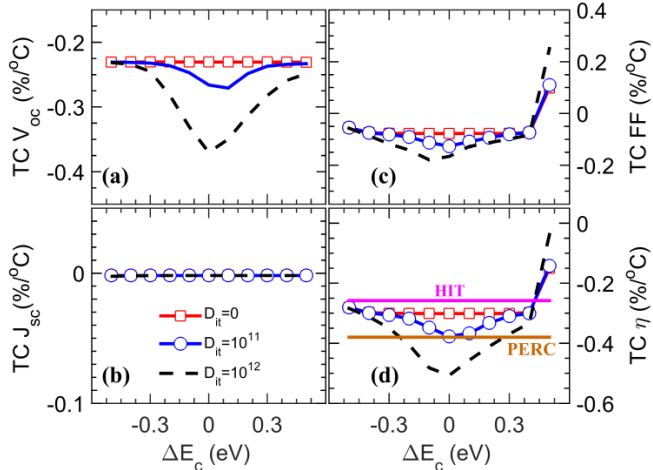

**Fig. 5**: The effect of the $\Delta E_c$ on temperature coefficient of parameters - (a) $V_{oc}$, (b) $J_{sc}$, (c) FF, and (d) efficiency for different $D_{it}$ (cm$^{-2}$eV$^{-1}$) values. The trends are broadly consistent with the analytical model for $V_{oc}$. FF follows the trends of $V_{oc}$ for low values of $\Delta E_c$ and for high values of $\Delta E_c$ it increases with temperature due to the increase in current at maximum power point.

The temperature coefficient of $J_{sc}$ (Fig. 5b) at small $\Delta E_c$ is zero due to (a) insignificant variation in net carrier generation rate, and (b) the over the barrier transport is unaffected for the range of $\Delta E_c$ under consideration (as predicted by the analytical model). The trends for the temperature coefficient of FF (Fig. 5c) are similar to temperature coefficient of $V_{oc}$ for most values of $\Delta E_c$. However, for large values of $\Delta E_c$, the current at maximum power point increases with temperature, which leads to an improvement in TCFF. Figure 5d shows the variation of temperature coefficient of efficiency which follows the trends of $V_{oc}$ and FF. The same figure also highlights a comparison of the temperature coefficient of CS solar cells with that of state of the art PERC and HIT solar cells[20]. It is evident that unless the band discontinuity is small and the interface quality is very bad, temperature coefficient of CS solar cells can be better than PERC solar cells (of course, the effect of HTL/Si interface should also be accounted for a proper comparison – which will be attempted in later sections). Further, HIT cells are almost universally better than both the CS as well as the PERC cells. However, we note that for very large $\Delta E_c$, CS solar cells show better TCη than even HIT cells. This is a regime where the charge collection efficiency and hence the FF is limited by over the barrier transport. So, although the TCη is better, the efficiency of such cells is still much lower than that of comparable HIT cells.

The previous section detailed the influence of minority carrier extraction interface on the temperature coefficients. Here, we extend the same to majority carrier extraction interface (i.e., c-Si/HTL). The effects of temperature and band discontinuity (i.e., $\Delta E_v$) at c-Si/HSL interface on the normalized solar cell performance metrics are shown in Fig.6. As before, here we assume ideal conditions at ESL/c-Si interface (i.e., zero band offset and no traps). While the trends in $V_{oc}$ degradation with temperature (Fig. 6a) are somewhat similar to the results for minority carrier collection (as discussed in Fig. 4a), the trends for $J_{sc}$ variation with temperature is distinctly different (i.e., compare Fig. 6b with Fig. 4b). Specifically, the results indicate a positive temperature coefficient for small $\Delta E_v$. As a result, the FF (Fig. 6c) and efficiency (Fig. 6d) trends are different compared to Fig. 4.

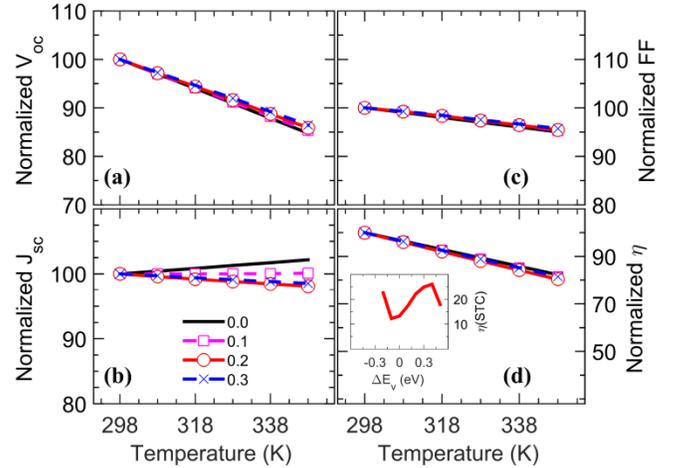

**Fig. 6**: The effect of the temperature on normalized performance metrics - (a) $V_{oc}$, (b) $J_{sc}$, (c) FF, and (d) efficiency of CS Si solar cells for $D_{it} = 10^{12}$ cm$^{-2}$eV$^{-1}$. The labels represent $\Delta E_v = 0$eV, $0.1$eV, $0.2$eV and $0.3$eV. The inset in (d) shows the effect of $\Delta E_v$ on efficiency with $D_{it} = 10^{12}$ cm$^{-2}$eV$^{-1}$.

The trends in $J_{sc}$ can be explained by using Fig. 7. Parts (a) and (b) of Fig. 7 indicate the E-B diagrams for $\Delta E_v = 0$ eV and for $\Delta E_v = 0.2$ eV, respectively. For these two cases, the variation in majority carrier density ($p_s$) and the net interface recombination are plotted in parts (c) and (d). Fig 7c shows that for $\Delta E_v = 0$ eV, the interface recombination remains unaffected by the temperature and that the $p_s$ increases with temperature. This improves the collection efficiency and hence leads to an increase in $J_{sc}$ with temperature as observed in Fig. 6c for $\Delta E_v = 0$ eV. However, Fig. 7d shows that the interface recombination increases with temperature and leads to a decrease in $p_s$ for $\Delta E_v = 0.2$ eV. This increase in interface recombination with temperature coupled with the decrease in $p_s$ contributes to a reduction in $J_{sc}$ and hence the temperature coefficient for $J_{sc}$ is negative at higher band discontinuities.

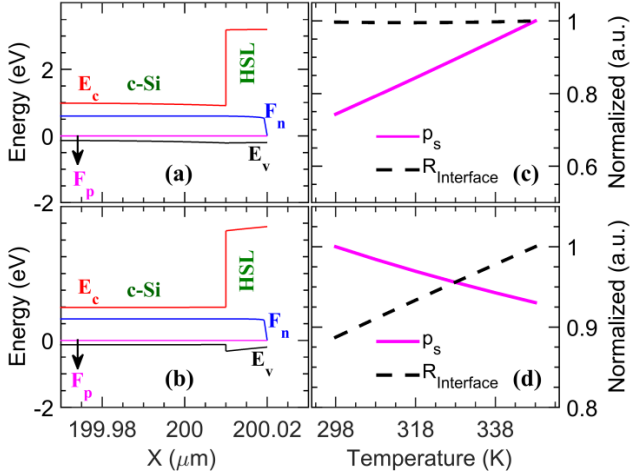

**Fig. 7**: The effect of $\Delta E_v$ on band bending and carrier densities at c-Si/HSL interface with $D_{it} = 10^{12} cm^{-2} eV^{-1}$ (numerical simulations, under illumination). Parts (a) and (b) show the energy band diagram for $\Delta E_v = 0eV$ and $\Delta E_v = 0.2eV$, respectively, at short circuit conditions. Parts (c, d) show the variation in normalized interface carrier density $p_s$ and normalized interface recombination with temperature at short circuit conditions for $\Delta E_v = 0eV$ and $\Delta E_v = 0.2eV$, respectively.

Figure 8 shows the effect of c-Si/HSL interface quality ($\Delta E_v$, $D_{it}$) on temperature coefficient of parameters for CS solar cell. As the band offset increases, interface recombination decreases and hence the $TCV_{oc}$ improves. These trends are similar to the results shown in Fig. 5a. However, the temperature coefficient for $J_{sc}$ (see Fig. 8b) shows significant variation at low values of $\Delta E_v$. The difference between the corresponding curves in Fig. 5b is due to lower band bending at the PP+ junction between HSL and Si for low values of $\Delta E_v$. Figure 8c and 8d show the variation in temperature coefficient of FF and efficiency respectively. We find that $TC\eta$ broadly follows the trends of $TCV_{oc}$. However, for small $\Delta E_v$, the trends are influenced by the $TCJ_{sc}$ as well. The comparison with state of the art PERC and HIT cells gives similar inferences as in Fig. 5d.

The combined effect of interface non-idealities at ESL/Si and HSL/Si interfaces on temperature coefficient is shown in Fig. 9. Here parts (a) and (b) show the variation in temperature coefficient when $\Delta E_c = \Delta E_v = 0eV$ and $\Delta E_c = \Delta E_v = 0.3eV$, respectively. For a given interface trap density, the lower band discontinuities result in a larger temperature coefficient. Curiously, the same results indicate that the quality of minority carrier extraction layer interface with Silicon is more critical for temperature coefficient of the device than the quality of majority carrier extraction layer interface with Silicon. Finally, we note that the effect of interface traps is not cumulative on the temperature coefficient, i.e., for same $D_{it}$ the sum of temperature coefficients with traps at only one interface (ESL/Si or HSL/Si) is more than the temperature coefficient with traps at both interfaces. Comparison with the state of the art cells indicates that while HIT cells may continue to boast excellent temperature coefficients, well designed CS solar cells might perform better than PERC cells.

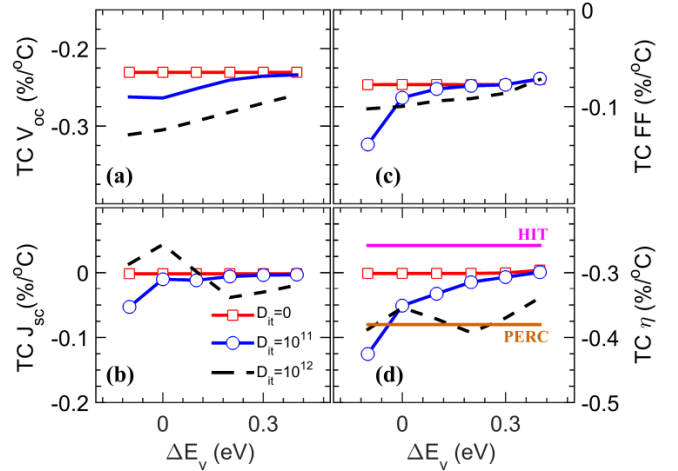

**Fig. 8**: The effect of $\Delta E_v$ on temperature coefficient of performance metrics - (a) $V_{oc}$, (b) $J_{sc}$, (c) FF, and (d) efficiency in the presence of interface traps. Due to the variation in $J_{sc}$ the curves for efficiency are different from the corresponding curves for $\Delta E_c$.

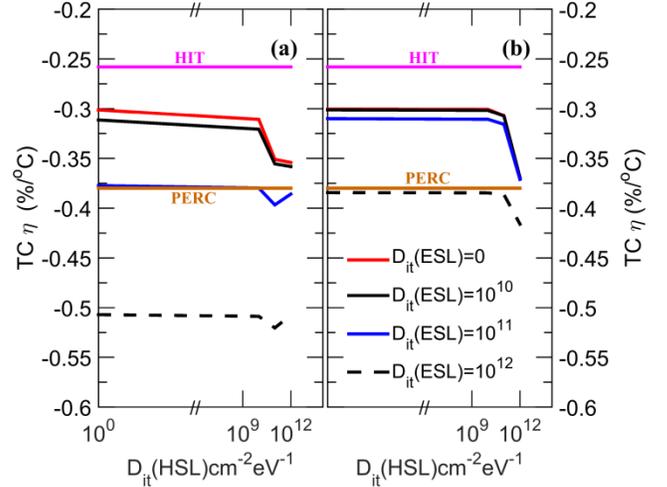

**Fig. 9**: The effect of traps at both the HSL and the ESL interface with c-Si on temperature coefficient for (a) $\Delta E_c = \Delta E_v = 0eV$ and (b) $\Delta E_c = \Delta E_v = 0.3eV$. The quality of minority carrier extraction layer interface with Silicon is more critical for temperature coefficient of the device than the quality of majority carrier extraction layer interface with Silicon.

## IV. CONCLUSIONS

To summarize, here we addressed the temperature coefficient of Si based CS layers as a function of the band discontinuity and interface quality. Through an analytical model we explored the functional dependence of temperature coefficient on such material and interface properties. These predications were validated using detailed numerical simulations. The results show that the temperature coefficient improves with band discontinuity if the interface quality is imperfect. Further for larger band discontinuity the temperature coefficient improves due to the increase in over the barrier transport with temperature irrespective of the interface quality. In addition, our results show that the passivation quality at the minority carrier layer interface is more critical than the majority layer interface and that the temperature coefficient is not cumulative, i.e., the temperature coefficient of a structure with traps at both interfaces is better than the sum of the temperature coefficient of separate device structures with traps being present at only one interface (either ESL/Si or HSL/Si). Finally our results indicate that with appropriate design, CS solar cells can achieve temperature coefficients better than PERC solar cells – although they might still be inferior to HIT solar cells. These interesting insights could be of broad interest to the community towards the optimization of process and material considerations for Si based CS solar cells.

## V. APPENDIX

**A. Parameters used in simulations:** The band offset $\Delta E_c$ at the ESL/c-Si interface is varied from $-0.6eV$ to $+0.6eV$, while the barrier for holes is kept fixed (2.28eV). Accordingly, in our simulations the ESL band gap varies from 2.8eV to 4eV, which is comparable to the band gap of $TiO_2$ (~3.4eV). At the c-Si/HSL interface, $\Delta E_v$ is varied from $-0.1eV$ to $+0.4eV$, while the barrier for electrons is kept fixed (2.28eV). Correspondingly, the HSL bandgap varies from 3.3eV to 3.8eV. For ease of analysis, we have used same dielectric constant (6.2) for both ESL and HSL. We consider uniform distribution of traps at the interface of CS layer and Si. The capture cross section of these traps was assumed as $10^{-16}cm^{-2}$. The rest of the parameters are provided in the table below.

| Parameter | c-Si | ESL | HSL |
|---|---|---|---|
| $N_c(cm^{-3})$ | $3.23 \times 10^{19}$ | $2.5 \times 10^{20}$ | $2.5 \times 10^{20}$ |
| $N_v(cm^{-3})$ | $1.83 \times 10^{19}$ | $2.5 \times 10^{20}$ | $2.5 \times 10^{20}$ |
| Mobility($cm^2V^{-1}s^{-1}$ (n, p)) | 1417, 470.5 | 20, 2 | 20, 2 |
| $\tau$ SRH (s) | $10^{-3}$ | $10^{-6}$ | $10^{-6}$ |
| Radiative Recombination coefficient ($cm^3s^{-1}$) [24] | $1.1 \times 10^{-14}$ | | |
| Auger Coefficients ($cm^6s^{-1}$) (n,p) [25] | $1 \times 10^{-31}$, $0.79 \times 10^{-31}$ | | |
| Doping($cm^{-3}$)n/p | p - $10^{17}$ | n - $10^{17}$ | p - $10^{17}$ |

**Table. 1**: Parameters used in numerical simulations.

**Acknowledgements:** The authors acknowledge Center of Excellence in Nanoelectronics (CEN) and National Center for Photovoltaic Research and Education (NCPRE), IIT Bombay for computational facilities.